\documentclass{article}
\usepackage[english]{babel}
\usepackage{amsfonts}
\usepackage{amsmath}
\usepackage{amssymb}
\usepackage{amsthm}
\usepackage{array}
\usepackage{multirow}
\usepackage{float}
\usepackage{booktabs}
\usepackage{color}
\usepackage{csquotes}

\usepackage{tikz}

\usepackage{authblk}
\usepackage{hyperref}

\newcommand{\PATH}[1]{\overline{#1}}
\newcommand{\bfa}{{\bf a}}
\newcommand{\bfb}{{\bf b}}

\newcommand{\bfe}{{\bf e}}

\newcommand{\bfu}{{\bf u}}
\newcommand{\bfv}{{\bf v}}

\newcommand{\bfx}{{\bf x}}
\newcommand{\bfI}{{\bf I}}

\newcommand{\bfP}{{\bf P}}
\newcommand{\bfQ}{{\bf Q}}
\newcommand{\bfT}{{\bf T}}
\newcommand{\bfX}{{\bf X}}
\newcommand{\ket}[1]{{\vert{#1}\rangle}}
\newcommand{\bra}[1]{{\langle{#1}\vert}}
\newcommand{\braket}[2]{{\langle{#1}\mid{#2}\rangle}}

\newcounter{productrules}
\newcounter{altrules}
\newcounter{tableone}

\begin{document}
\title{The symmetrical foundation of Measure, Probability and Quantum theories \thanks{J. Skilling, K. H. Knuth, 2018. The symmetrical foundation of measure, probability and quantum theories, \textit{Annalen der Physik}, 1800057. \url{https://doi.org/10.1002/andp.201800057}}}

\author[1]{\small John Skilling}
\author[2]{\small Kevin H. Knuth}

\affil[1]{\footnotesize Maximum Entropy Data Consultants Ltd, Kenmare, Ireland}
\affil[2]{\footnotesize Department of Physics, University at Albany (SUNY), Albany NY, USA}


\maketitle

\abstract{
Quantification starts with sum and product rules that express combination and partition.
These rules rest on elementary symmetries that have wide applicability, which explains why arithmetical adding up and splitting into proportions are ubiquitous.
Specifically, measure theory formalises addition, and probability theory formalises inference in terms of proportions.

Quantum theory rests on the same simple symmetries, but is formalised in two dimensions, not just one, in order to track an object through its binary interactions with other objects.
The symmetries still require sum and product rules (here known as the Feynman rules), but they apply to complex numbers instead of real scalars,
 with observable probabilities being modulus-squared (known as the Born rule).
The standard quantum formalism follows.
There is no mystery or weirdness, just ordinary probabilistic inference.
}


\section*{Contents:}

\noindent 1. Introduction

\noindent 2. Measures

\noindent 3. Probability

\noindent\phantom{space} 3.1. The sum rule				\\
\noindent\phantom{space} 3.2. The product rule			\\
\noindent\phantom{space} 3.3. Proportions				\\
\noindent\phantom{space} 3.4. Bayes' theorem				\\
\noindent\phantom{space} 3.5. Interpretation of probability		\\
\noindent\phantom{space} 3.6. Usage of probability

\noindent 4. Quantum theory

\noindent\phantom{space} 4.1. The product rules			\\
\noindent\phantom{space} 4.2. Modulus and phase			\\
\noindent\phantom{space} 4.3. The sum rules				\\
\noindent\phantom{space} 4.4. Probability assignments		\\
\noindent\phantom{space} 4.5. Hilbert space				\\
\noindent\phantom{space} 4.6. Single object

\noindent 5. Commentary

\section{Introduction}

``\emph{Quantum mechanics will cease to look puzzling only when we will be able to derive the formalism of the theory from a set of simple assertions about the world.}''
 --- Carlo Rovelli \cite{Rovelli:FQXI2013}
\medskip

Our job in science is to make sense of our observations.
This is very general, and we seek corresponding clarity and simplicity.
General theory must apply to all cases, and our strategy of \emph{eliminative induction} \cite{Caticha:2009:rational-belief}
 is to exclude theories which give ``wrong'' results in particular cases, until there remains just a single candidate theory which we can then recommend with confidence.

Symmetries are particularly valuable tools for eliminating ``wrong'' behaviour \cite{Cox:1946,Caticha:1998,GKS:PRA,GK:Symmetry,Knuth+Skilling:2012}.
If $A$ is supposed to equal $B$, then all the other possibilities with $A \ne B$ are immediately excluded.
Symmetries are particularly powerful when applied to simple systems because ``wrong'' behaviour is seen most clearly there, where judgement is least subjective.
Accordingly, we proceed by considering simple thought-experiments, whose behaviour should be uncontroversial.

In classical physics, objects can be detected passively, unchanged by interaction with a probe.
That's a valid limit for when the object dominates the probe, and its quantification leads to standard scalar measure theory (``stuff adds up'').
More fundamentally, though, a probe is a partner object, not necessarily so small as to to be dominated.
But, if our object can perturb a partner object, then by symmetry the partner object can also perturb our object.
We could assign either r\^ole to either.

Our calculus, whatever it is, must be capable of representing such interactions.
The strategy of eliminative induction leads us to exclude all calculi that fail particular tests of acceptable interaction.
This insight that interactions are basic is the source of ``quantum-ness''.
It leads to the \emph{Feynman rules} \cite{Feynman:1948},
 which represent elementary interactions by \emph{pairs} of numbers which intertwine according to the rules of complex arithmetic,
 and to the \emph{Born rule} \cite{Born:1926} which relates those complex numbers to outcomes.

By this stage, we have reduced what could have been arbitrary rules to a single defined calculus.
Hilbert space can then be constructed \cite{vonNeumann:1996(1955)} and the rest of quantum theory follows.
The quantum rules are universally applicable to all of physics, though of course classical limits are adequate for appropriately large-scale objects.

In this paper, we use the same thought-experiments and the same symmetries for both classical and quantum situations.
This gives a straightforward unified derivation of measure theory, probability theory, and quantum theory.
Two formal operations are uncovered -- a sum rule and a product rule.
Measures formalise quantitation (kilograms, coulombs and so on).
Probability formalises our inferences.
Quantum theory tracks objects through their interactions.

\section{Measures}

``\emph{I have tried, with little success, to get some of my friends to understand my amazement that the abstraction of integers for counting is both possible and useful. \dots
To me, this is one of the strongest examples of the unreasonable effectiveness of mathematics.
Indeed, I find it both strange and inexplicable.}'' --- Richard Hamming \cite{Hamming:1980}.
\medskip

Starting at the beginning with minimalist foundation, we  might, perhaps, think of a shopping basket of fruit --- apples, bananas and so on.
Consider the operator $\oplus$ which combines disjoint objects $A$ and $B$ into a composite object $A\oplus B$.
We list some basic properties of $\oplus$ which are commonly applicable.

\begin{description}
\item[$\boxed{Closure}$]{\quad \begin{tabular}{l} A combination of objects is an object \\[-2pt] in its own right. \end{tabular}
\begin{equation}
    \text{$(A \oplus B)$ is an object, where $A$ and $B$ are disjoint}
\end{equation}
}
\item[$\boxed{Commutative}$]{\quad \begin{tabular}{l} The order of objects \\[-2pt] does not matter. \end{tabular}
\begin{equation}
	A \oplus B = B \oplus A
\end{equation}
}
\item[$\boxed{Associative}$]{\quad \begin{tabular}{l} The order of combination \\[-2pt] does not matter. \end{tabular}
\begin{equation}
	(A \oplus B) \oplus C = A \oplus (B \oplus C)
\end{equation}
Commutativity and associativity together mean that objects can be arbitrarily shuffled.
\vskip4pt
}
\item[\boxed{\text{\it Limitless}}] {\quad \begin{tabular}{l} Equivalent, but disjoint, objects can \\[-2pt] be combined without restriction. \end{tabular}
\begin{equation}
    \begin{array}{r} \text{$A,\, 2A,\, 3A,\ldots\ldots$ are all different,} \\ \text{where $(n\mathord+1)A = (nA) \oplus A$} \end{array}
\end{equation}
Here individual $A$'s are equivalent but disjoint objects $A_1,A_2,A_3,\ldots\,$.
We intend no infinite limit here.
All we claim is the freedom to include more objects, limited only by our resources and patience, but not somehow limited by an intellectual barrier.
If there is a cardinality restriction, we will never reach it.
It would have no practical consequences for us,  so we are free to ignore it.
}
\end{description}

If these properties (closure, commutativity, associativity,  limitless) are all accepted, then
\begin{equation}  \label{eq:generalsum}
	\begin{array}{ll} A \oplus B & \text{\quad is represented (up to isomorphism)} \\
	 	& \text{\quad by the component-wise sum \quad $\bfa + \bfb$} \end{array}
\end{equation}
where $\bfa = (a_1,a_2,\ldots,a_n)$ and $\bfb = (b_1,b_2,\ldots,b_n)$ represent $A$ and $B$ by $n$-tuples of scalar elements.
This is the general \emph{sum rule}.
In accordance with practical feasibility, the rule is demonstrated by construction \cite[appendix A]{Knuth+Skilling:2012},
 successively incorporating new objects and introducing a new dimension whenever a new object is not commensurable with the existing set,
 commensurability of $X$ and $Y$ being defined in the usual way as the existence of non-zero integers $m$ and $n$ such that $m$ of $X$ can be deemed equivalent to $n$ of~$Y$.

Equivalence requires equality of representation and non-equivalence requires inequality.
The quoted sum rule preserves all such correspondences.

According to the construction, the elements $a$ and $b$ start as integers, which can then be rescaled for convenience to rational numbers,
 whence they become observationally indistinguishable from general real numbers.
This construction is finite: macroscale continuity emerges for arbitrarily extensive systems but we do not assume it.

Like all basic assignments, commensurability is decided by practical judgement.
Thus $\pi$ (mathematically known to be irrational) and its accurate approximation 3.14159265358979 would usually be deemed equivalent for practical purposes,
 allowing circumferential lengths to be commensurate with diameter lengths.
The mathematical distinction between rational and real numbers has zero practical impact, so can always be ignored for rational inference in practical science.

The representation is only forced ``up to isomorphism'', meaning that any 1:1 re-grading  is logically equivalent and has exactly the same analytical power.
Conversely, a representation that was not in 1:1 correspondence would break some of the equivalences and/or non-equivalences, so would be rejected.
The convenience of ``+''  operating arithmetically  is so great that we adopt it as a near-universal convention.
For one thing, its compact connected topology fits naturally with our intuitive notion of locality: small changes have small effect.

Within the additive  $n$-tuple formulation, the only freedom consistent with preservation of ``+'' and the original commensurabilities is linear shear $\bfx' = \bfT\bfx$
 by an arbitrary non-singular (affine) $n\mathord\times n$ transformation matrix~$\bfT$.

\begin{description}
\item[$\boxed{Dimension}$]{\quad If our objects are fully commensurable, so that there is only one relevant property (as when a shopping basket holds apples only),
 the dimension shrinks to 1 and the sum rule takes scalar form.
\begin{equation}
	\text{\quad$A \oplus B$  \qquad is represented by \qquad $a + b$\quad}
\end{equation}
}
\end{description}   
Such a quantity is known in physics as an extensive variable.
The only freedom consistent with preservation of ``+'' and the original commensurabilities is linear rescaling to different units.
The quantity may or may not be signed.
Electric charge, for example, is signed, but mass is not.
A single-signed quantity, positive by convention, is known in mathematics as a \emph{measure}.

Familiarity makes addition seem obvious, indeed trivial.
Here, though, we see \emph{why} additivity is so ubiquitous \cite{Knuth:FQXI2015}.
It's required by elementary symmetries that are commonly upheld.

To see whether additivity is required, all we have to do is check the boxes.
If they are satisfied, then the sum rule must apply, in the appropriate dimension.
We do not need bespoke derivations for each application.
Neither do we need sophisticated formalism for such simple requirements.


\section{Probability}

``\emph{Probability is expectation founded upon partial knowledge.}'' --- George Boole \cite{Boole:1854}.
\medskip

\addtocounter{equation}{1}
\setcounter{tableone}{\theequation}
The operation inverse to combination is \emph{partition}, in which a composite object is progressively decomposed,
 if necessary all the way down to a notional substrate  (\theequation) of a-priori-equivalent  base states.
\begin{equation*}
\hbox{
\begin{tabular}{|c|c|c|c|c|c|c|c|c|lll}
\cline{1-9}
	 \multicolumn{9}{|c|}{$O$}														&\ Object	 		& $O$			&	\\
\cline{1-9}
	 \multicolumn{6}{|c|}{$A$}								& \multicolumn{3}{c|}{$D$}	&\ Partition 		& $O = A \oplus D$	&	\\
\cline{1-9}
	 \multicolumn{2}{|c|}{$B$}		& \multicolumn{4}{c|}{$C$}	& \multicolumn{3}{c|}{$D$}	&\ Partition 		& $A = B \oplus C$	&	\\
\cline{1-9}
	 \multicolumn{2}{|c|}{$B$}		& \multicolumn{7}{c|}{$E$}								&\ Combination 	& $C \oplus D = E$	&	\\
\cline{1-9}
	 \multicolumn{9}{|c|}{$O$}														&\ Combination 	& $B \oplus E = O$	&	\\
\cline{1-9}
	 \!\!$\bullet$\!\!	& \!\!$\bullet$\!\!	& \!\!$\bullet$\!\!	& \!\!$\bullet$\!\!	& \!\!$\bullet$\!\!	& \!\!$\bullet$\!\!	& \!\!$\bullet$\!\!	& \!\!$\bullet$\!\!	& \!\!$\bullet$\!\!	&\multicolumn{2}{l}{\ Notional substrate}  			& \!\!(\theequation) \\
\cline{1-9}
\end{tabular}
}
\end{equation*}
We always have the objects that interest us, but the substrate might be so arbitrarily deep that we never bother to use it.
In this way, we keep matters simple by avoiding any technical assumptions about infinitesimal limits.

\subsection{The sum rule}

Partitioning by itself has the structure of a tree, more specifically a directed rooted tree sourced at the root, with  nodes that can be combined together and joined by source-to-destination paths such as $\PATH{BA}$ from source $A$ to destination $B$.
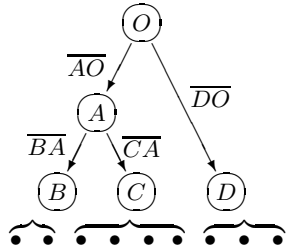
\begin{center}
\begin{picture}(84,93)(0,0) 
\put(-58,0){
\put(100,7){\makebox(0,0){$\overbrace{\bullet\ \ \,\bullet}\ \ \overbrace{\bullet\ \ \bullet\ \ \bullet\ \ \bullet}\ \ \overbrace{\bullet\ \ \bullet\ \ \bullet}$}}
\put(66,22){\makebox(0,0){$B$}}		\put(66,22){\oval(14,14)}
\put(96,22){\makebox(0,0){$C$}}		\put(96,22){\oval(14,14)}
\put(130,22){\makebox(0,0){$D$}}		\put(130,22){\oval(14,14)}
\put(81,52){\makebox(0,0){$A$}}		\put(81,52){\oval(14,14)}
\put(98,86){\makebox(0,0){$O$}}		\put(98,86){\oval(14,14)}
\put(77,44){\vector(-1,-2){7}}			\put(85,44){\vector(1,-2){7}}
\put(62,40){\makebox(0,0){$\PATH{BA}$}}	
\put(98,39){\makebox(0,0){$\PATH{CA}$}}
\put(94,78){\vector(-1,-2){9}}			\put(102,78){\vector(1,-2){24}}
\put(77,70){\makebox(0,0){$\PATH{AO}$}}	\put(124,58){\makebox(0,0){$\PATH{DO}$}}
}
\end{picture}
\end{center}

Quantification of the partitions obeys the symmetries of measure and hence (with dimension 1 here) we have the scalar \emph{sum rule}
\begin{equation} \label{eq:sum}
    p(B \oplus C) = p(B) + p(C)
\end{equation}
for the disjoint destinations  $B$ and  $C$ from a binary source node  $A = B\oplus C$.
The function $p(\cdot)$ returns the scalar valuation of its argument (here a node of the tree).

Quantification is thus additive over whatever values are assigned to the substrate (if we bother to construct it).
Using unit weights in the example above, $B$ would be quantified as 2, $C$ as 4, $A = B\oplus C$ as $6 = 2+4$ and so on.

\subsection{The product rule}

We are also interested in the quantification  of paths themselves.

Paths can be chained  by concatenation $\circ$, so that $\PATH{UV} \circ \PATH{VW}$ is also a path because $\PATH{UV}$ starts where $\PATH{VW}$ ends.
In the example above, paths  $\PATH{BA}$ and $\PATH{AO}$ chain to form $\PATH{BA} \circ \PATH{AO} = \PATH{BO}$.
Chaining is formalised as closure of $\circ$.
\begin{equation}
	\boxed{\text{\it Closure of\ \ $\circ$}} \enskip \text{$\PATH{UW} =  \PATH{UV} \circ \PATH{VW}$ is also a path}
\end{equation}
%


We also require right-distributivity.
In the example, $A$ was $B\oplus C$ regardless of the provenance of $A$, whether that was just $A$ itself or $O$ or in fact any object $X$ that covered $A$.
In terms of source-to-destination paths, the destination relationship $A=B\oplus C$ reflects combination of substrate parts, which continues unchanged for any covering source.

This is what enables us to carve out pieces of the Universe for local investigation, independent of what may be happening elsewhere.
We could not do science without it.
Although the idealisation may never be totally attained in practice, we require our calculus to \emph{allow} analysis of arbitrarily isolated systems.
This is formalised as
\begin{equation}
\boxed{\text{\it Right-distributive $\circ$}} \enskip  \begin{array}{l} (\PATH{UY} \oplus \PATH{VY}) \circ \PATH{YZ} = \\
											\quad (\PATH{UY} \circ \PATH{YZ}) \oplus (\PATH{VY} \circ \PATH{YZ}) \end{array}
\end{equation}

\noindent Right-distributivity implies that the quantification $p(\PATH{XZ})$ of a $Z$-to-$X$ source-to-destination
 path is linear in the destination quantity $p(X)$, with the only remaining freedom being a scale factor which may depend on the source~$Z$.
\begin{equation} \label{eq:pqXZ}
	p(\PATH{XZ}) = p(X) f(Z)
\end{equation}

The scaling $f$ is some as-yet-unknown function.
Likewise for other arguments:
\begin{equation}
	p(\PATH{XY}) = p(X) f(Y)\,,\qquad p(\PATH{YZ}) = p(Y) f(Z)\,.
\end{equation}
The representation of  $\PATH{UV}\circ \PATH{VW}$  is to be constructed from the representations of  $\PATH{UV}$ and $\PATH{VW}$, so
\begin{equation} \label{eq:phi}
	 p(\PATH{UV}\circ \PATH{VW}) = \phi\big( p(\PATH{UV}) , p(\PATH{VW}) \big)
\end{equation}
for some function $\phi$ representing $\circ$.
But  $\PATH{UV}$ and $\PATH{VW}$  chain to become the path  $\PATH{UW}$ with source $W$ and destination $U$  so, using \eqref{eq:pqXZ}--\eqref{eq:phi},
\begin{equation} \label{eq:qfqf}
	 p(U) f(W) = \phi\big(\, p(U) f(V) ,\, p(V) f(W) \,\big)
\end{equation}
The left side being linear in  $p(U)$ and in $f(W)$, the right side must be also.
Hence $\phi$ is bilinear, $\phi(a,b) = \gamma  ab$ with some coefficient $\gamma$.

On setting $\gamma=1$ by choice of scale factor, equation \eqref{eq:phi} reads
\begin{equation} \label{eq:prod}
	 p(\PATH{UV}\circ \PATH{VW}) = \gamma p(\PATH{UV})\, p(\PATH{VW})  \qquad\text{with $\gamma=1$}
\end{equation}
which is the \emph{product rule}.

\subsection{Proportions}


The imposed symmetries require relationships along paths to chain multiplicatively.
Consequently, node quantities all scale according to the magnitude of the root of the partition tree ($p(O)$ in the example above).
Thus
\begin{equation} \label{eq:scaling}
	p(B) = p(\PATH{BA})\, p(\PATH{AO})\, p(O)  
\end{equation}
and so on.
We also get scale-independent relationships
\begin{equation} \label{eq:Laplace}
	p(\text{\tt dest}\circ \text{\tt source}) = \frac{p(\text{\tt dest})}{p(\text{\tt source})}
\end{equation}
as ratios of quantities, so that the scaling function $f$ was $1/p$.
In the example above, we have $p(\PATH{AO})=6/9$, $p(\PATH{BA})=2/6$ and so on,
 and we could equally well have written the sum rule \eqref{eq:sum} in terms of binary-tree paths as
\begin{equation}
     p(\PATH{BA} \oplus \PATH{CA}) = p(\PATH{BA}) + p(\PATH{CA})
\end{equation}

Although this behaviour is ``obvious'', and it would be difficult to think of a plausible competitor,
 we will use our deeper justification from symmetries in more sophisticated setting when we come to quantum theory.

In this derivation, associativity is an emergent property of the representation.
However, associativity
\begin{equation} \label{eq:assoc}
\boxed{\text{\it Associative\ $\circ$}} \ \ \   (\PATH{UV} \circ \PATH{VW}) \circ\PATH{WX} = \PATH{UV} \circ (\PATH{VW} \circ \PATH{WX})
\end{equation}
of chained paths themselves is  itself  an intuitive requirement that could be taken as axiomatic \cite{Knuth+Skilling:2012}.
Indeed, there is a nexus of interrelated ``obvious'' properties, where the selection of which is axiomatic and which is emergent is to some extent a matter of choice.

\subsection{Bayes' theorem}

So far, the derivation has done no more than select simple proportion as the uniquely favoured relationship between partitioned measures.
However, probability obeys the same selected symmetries, so is an example of a partitioned measure  necessarily obeying the same calculus.
It is slightly special in that the root of the partition tree is assigned dominant status as the ``provisional truth'' from which subsidiary values derive.
This magnitude $p(O)$ is usually assigned as 1, or 100 if percentages are being used.

Partitioning and combination can be filled out to a Boolean lattice defined by  join $\vee$ (logical OR) which upgrades $\oplus$ to non-disjoint components,
 and meet $\wedge$ (logical AND) which identifies overlaps \cite{Knuth:measuring,Knuth:Annalen-der-Physik}.
In this context,  we recognise $p(\PATH{BA})$ as the transition probability from source $A$ to destination $B$ and traditionally written as $\Pr(B\mid A)$.
Symmetry of the product rule then gives \emph{Bayes' theorem}
\begin{equation} \label{eq:Bayes}
	\underbrace{{\Pr}(\theta)}_\text{\it Prior}\, \underbrace{{\Pr}(D\mid\theta)}_\text{\it Likelihood} =
	 \underbrace{{\Pr}(D)}_\text{\it Evidence}\, \underbrace{{\Pr}(\theta\mid D)}_\text{\it Posterior} \qquad\parallel M
\end{equation}
for the inference of parameters $\theta$ (``the source'') from data $D$ (``the destination'') --- in the context of model $M$
 so that all probabilities are additionally conditional on that hypothesised background information.

\subsection{Interpretation of probability}

Awkwardly, the traditional but idiosyncratic solidus notation suggests that there's something special about \hbox{$\Pr(\cdot\mid\cdot)$}.
There isn't.
Probability just obeys the same simple laws of proportion that apply widely elsewhere.
It needs no bespoke inference-specific derivation, though several different ones \cite{Cox:1946,Kolmogorov:1950,DeFinetti:1931,Jaynes:Book} have their devotees.

A calculus does not fix a context.
It applies to any application that obeys its requirements.
Probabilities \emph{are} proportions (of provisional truth) in just the same way as sectors of a pie chart \emph{are} proportions (of a disc).
The two are isomorphic to each other and to proportions, and are thereby interchangeable.
We may choose to represent an abstract probability of $\frac{1}{12}$ by a displayed $30^\circ$ sector of a disc but the change makes no difference to our manipulations or conclusions.

Philosophers make a distinction between the ontological sector of a plum pudding (which exists physically)
 and the epistemological probability (mental belief or expectation) that the traditional sixpenny coin will be found in that particular sector.
But the two are quantified identically.
They are isomorphic.

The ontology-epistemology divide is, for quantitation at least, a distinction without a difference.
A bit of information carries no flag to inform us whether it was assigned by a conscious agent or by a mechanical switch.
Our job in science is to make sense of our observations, not to indulge in  empty  disputation between isomorphic views.

Our goal here is derivation of a calculus fit for general purpose.
Ontology and epistemology share the same symmetries, the same rules, and the same assignments.
So they share a common calculus.

\subsection{Usage  of probability}

Bayes' theorem is acknowledged as the foundation of rational inference.
It allows us to infer parameters from data within the context of stated background assumptions.

Although the prior distribution is in general arbitrary, it must conform to any symmetry of the application because equivalent situations should be modelled equivalently.
Thus, if the application imposes left-right symmetry of $\theta$ across the origin, the prior $\Pr(\theta)$ should be an even function.
Rotational invariance leads to a prior independent of angle.
And so on.

In summary, a prior for the controlling variables  must be assigned and it must conform to any symmetry of the application.
Those are our requirements as we now turn to the language of quantum theory.

\section{Quantum Theory}

``\emph{This theoretical failure to find a plausible alternative to quantum mechanics, even more than the precise experimental verification of linearity,
 suggests to me that quantum mechanics is the way it is because any small change in quantum mechanics would lead to logical absurdities.}'' --- Steven Weinberg \cite{Weinberg:1992:dreams}
\medskip

We start with an identifiable object.
Its only property of relevance will be its existence in a particular target state.
More precisely, since we will be dealing with uncertainty, we probe the \emph{probability} of  its  existence in the target state.

Probabilities become observable as long-term success ratios \eqref{eq:Laplace}, so we posit repeated independent experiments in which the object is either in the target state or not.
We simplify this analysis by letting the target states appear sporadically in essentially-continuous time.
The number of instants when the target state does not appear is then huge and essentially unaffected by the limited number of successes, so absences can be ignored.
Targets  are supplied  at some mean Poisson rate --- a single scalar number --- and our object need not even exist between those events.

A \emph{probe} is a partner object through which we observe the target by interacting with it.
Our knowledge of the target comes from such interactions, so the fundamental calculus we need is one of \emph{interactions}, most simply between just two objects, target and probe.

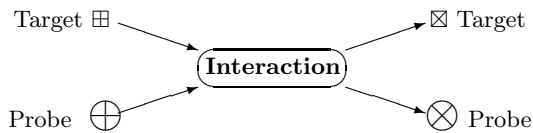
\begin{center}
\begin{picture}(200,46)(0,0) 
\put(0,-28){
\put(100.5,51){\makebox(0,0){\bf Interaction}}	\put(100,50){\oval(56,14)}
\put(128,57){\vector(3,1){30}}	\put(178,68){\makebox(0,0){$\boxtimes$ Target}}
\put(128,43){\vector(3,-1){30}}	\put(178,32){\makebox(0,0){{\LARGE$\otimes$} Probe}}
\put(42,67){\vector(3,-1){30}}	\put(22,68){\makebox(0,0){Target $\boxplus$ }}
\put(42,33){\vector(3,1){30}}	\put(22,32){\makebox(0,0){Probe {\LARGE$\,\oplus$}}}
}
\end{picture}
\end{center}

The target is detected by a change of state in the probe,
 which may optionally be amplified into a potentially permanent readable representation such as symbols on a printed page or bits in a classical computer.
Conceivably, amplifiers obeying the quantum rules might be provably impossible.
We do not pursue that nihilistic possibility, pointing instead to demonstrable existence of amplifiers that do obey the rules, thereby enabling measurement.

Since a basic interaction involves two objects, here target and probe, it is natural to suppose that its fundamental representation involves \emph{two} numbers, not just one.
Since we have no way of ``seeing'' inside an interaction, we may not assume that these two real numbers relate separately to target and probe.
We formalise this key assumption as the
$$
	\boxed{\begin{array}{ll}
		\text{\emph{Pair Postulate:}} 	& \text{Interactions are represented by} 			\\[-2pt]
								& \text{intertwined \emph{pairs} of real numbers.}	\end{array}}
$$
Bold-face symbols ``$\bfu,\bfv,\dots$'' represent the pairs associated with particular paths that our target may traverse.
Quantification will be through scalar observation, so we associate a scalar $p(\bfu)$ with each pair $\bfu$.

Looking ahead, pairs will behave as complex numbers which will combine into more general quantum wave functions quantified by modulus-squared.

\subsection{The product rules}

Paths obey the symmetries listed above, to which their representations must conform.
Accordingly, their associated scalars obey the standard product rule \eqref{eq:prod}
\begin{equation} \label{eq:scalarprod}
  	p(\bfu \mathop\circ \bfv) = p(\bfu) p(\bfv)
\end{equation}
with corollary that scalar valuations at  nodes in a partition tree  scale to the common root valuation as in \eqref{eq:scaling}.

The associated pair product rule, promoted from \eqref{eq:prod} into two dimensions, is
\begin{equation} \label{eq:gammaprod}
  	(\bfu \mathop\circ \bfv)_i = \sum_{jk}\gamma_{i\!j\!k}u_j v_k
\end{equation}
Awkwardly, there's a lack of specificity here.
The dimension being two, there are eight constant coefficients $\gamma_{...}$, not just one.
Their arbitrariness can be reduced by applying appropriate linear shear (no longer just a single scale which can immediately be set to 1) to the pairs.

However, a $2\times2$ shear matrix has only 4 components, which are insufficient to reduce 8 $\gamma$'s to a single standard form.
To resolve this, we adopt associativity of product \eqref{eq:assoc} as an additional requirement.
Chaining would be naturally associative in any dimension but now, in two, the property is needed.

We have shown \cite{GKS:PRA}
 that associativity reduces the bilinear product rule \eqref{eq:gammaprod} to three different classes, each of whose coordinate axes can be sheared
  into a standard form (\!\cite[eq. 20]{GKS:PRA} with discriminant $\mu = -1,0,1$ respectively).
\addtocounter{equation}{1}
\setcounter{productrules}{\theequation}
\begin{equation*}
	\begin{array}{ll} \bfu \circ \bfv  =	&  \displaystyle\binom{u_1v_1 - u_2v_2}{u_1v_2 + u_2v_1} 															\\
							&  \text{\ \ or\ \ }   \displaystyle\binom{u_1v_1}{u_1v_2 + u_2v_1} \text{\ \ or\ \ }  \displaystyle\binom{u_1v_1 + u_2v_2}{u_1v_2 + u_2v_1}	\end{array}
	\hfill{(\theequation\text{a,b,c})}
\end{equation*}
The other two apparently-allowable classes \cite[eqs. 22 and 23]{GKS:PRA}
\addtocounter{equation}{1}
\begin{equation*}
    \bfu \circ \bfv =     \binom{u_1v_1}{u_2v_1}	\text{\ \ or\ \ }  \binom{u_1v_1}{u_1v_2} 		\hfill{(\theequation\text{a,b})}
\end{equation*}
are  degenerate and quickly rejected because one of the factors (either $\bfu$ or $\bfv$) is only present with one component, so it operates as a scalar, contrary to the pair postulate.
So (apparently) pairs have three alternative product rules ({\theproductrules}a,b,c).

These three product rules can conveniently be cast in the common form of multiplicative moduli and additive phases
\begin{equation} \label{eq:modphase}
\begin{array}{c}
	\log|\bfu\circ\bfv| = \log |\bfu| + \log |\bfv|	\\
	\arg(\bfu\circ\bfv) = \arg(\bfu) + \arg(\bfv)	\end{array}
\end{equation}
by defining modulus and phase in a or b or c respectively as
\addtocounter{equation}{1}
\setcounter{altrules}{\theequation}
\begin{equation*}
	\begin{array}{lll} 				& |\bfx|  = \sqrt{x_1^2 + x_2^2}\,,	&\quad \arg\,\bfx = \arctan(x_2/x_1) 		\\
				 \text{or\quad} 		& |\bfx|  = |x_1|\,,				&\quad \arg\,\bfx = x_2/|x_1| 			\\
				 \text{or\quad} 		& |\bfx|  = \sqrt{x_1^2 - x_2^2}\,,	&\quad \arg\,\bfx = \rm{arctanh}(x_2/x_1)	\end{array}		\hfill{(\theequation\text{a,b,c})}
\end{equation*}
Definition ({\thealtrules}a) is standard polar coordinates, ({\thealtrules}c) is the hyperbolic analogue, and ({\thealtrules}b) is the common limit (rescaled) for dominant first component.

\subsection{Modulus and phase}

Assignments on paths can be chained back up to the root node of a partition tree, as in \eqref{eq:scaling}.
There, target states are supplied at an observable mean Poisson rate.
That will constrain the scalar valuation $p$ which is itself a function $p(\bfu)$ of its underlying pair.
We now proceed to determine this function by requiring consistency between the additive and multiplicative behaviour of scalars, and of pairs.

Three quantities, logarithmic scalar valuation from \eqref{eq:scalarprod} and log-modulus and phase from \eqref{eq:modphase}, all behave linearly under chaining.
Hence $\log p$ is incremented linearly in log-modulus, and in phase, when an extra path is concatenated.
Accordingly, the relationship is linear:
\begin{equation}
	\log p(\bfx) = \alpha \log|\bfx| + \beta\arg(\bfx)
\end{equation}
for some constants $\alpha$ and $\beta$.
This is the part of the pair $\bfx$ that is constrained by the observations which yield $p$ as a mean rate.
The conjugate variable, independent of $p$ so not constrained by observation, is
\begin{equation}
	\xi = \beta \log|\bfx| - \alpha\arg(\bfx)
\end{equation}
Its origin can be offset arbitrarily without observational effect, implying that it is to be ascribed a uniform prior probability distributrion $\Pr(\xi) = \text{constant}$.

But the range of $\xi$ is in general infinite, making the constant zero so that every finite range has zero probability.
This is unusable: ``improper'' priors have no place in rational inference.
The paradox is overcome only by setting $\beta=0$ and selecting the first multiplicative alternative in which $\arg\,\bfx$ takes $2\pi$-periodic wraparound form.

Phase can now be considered limited to $0 \le \arg\,\bfx < 2\pi$, uniformly distributed as
\begin{equation}
	\Pr(\arg\,\bfx) = 1/2\pi
\end{equation}
Meanwhile, $\beta=0$ so that \cite{Tikochinsky:1988}
\begin{equation} \label{eq:alpha}
	p(\bfx) = |\bfx|^\alpha
\end{equation}
where $|\bfx|$ is the standard complex-number modulus ({\thealtrules}a).
Only $\alpha$ now needs to be determined.

\subsection{The sum rules}

Under partition or combination, pairs combine component-wise according to \eqref{eq:generalsum} as
\begin{equation} \label{eq:psum}
	\bfu\oplus\bfv = \binom{u_1 + v_1}{u_2 + v_2}
\end{equation}
\emph{Pairs are now recognised as complex numbers}, because theirs are the relevant addition and multiplication rules \eqref{eq:psum} and ({\theproductrules}a).
In quantum terminology, these are the \emph{Feynman rules}.

Consider the combination of two independently-sourced unit input rates.
These are underlain by unit-modulus pairs
\begin{equation}
	\bfx_1 = e^{i\theta}\,,\qquad \bfx_2 = e^{i\phi}
\end{equation}
where $\theta$ and $\phi$ are uniformly distributed random phases.
Upon combination, their sum $\bfX = \bfx_1 + \bfx_2$ has scalar valuation
\begin{equation}
	P = p(\bfX) = |\bfx_1 + \bfx_2|^\alpha
\end{equation}
whose observable mean rate
\begin{equation}
	\big\langle P \big\rangle = \left\langle \big\vert e^{i\theta} + e^{i\phi}\big\vert^\alpha \right\rangle_{\theta,\phi} = \frac{\Gamma(\alpha+1)}{\Gamma(\frac{\alpha}{2}+1)^{\,2}}
\end{equation}
must equal 2 (the sum of the two unit inputs).
The solution of this equation is $\alpha=2$, so that
\begin{equation} \label{eq:Born}
	p(\bfx) = |\bfx|^2
\end{equation}
which is the \emph{Born rule}.


\subsection{Probability assignments}

With pairs $\bfx$ identified as complex numbers,
 their calculus identified as their addition and multiplication (the Feynman rules \eqref{eq:psum} and ({\theproductrules}a)),
 and their relationship to observed rates $p(\bfx)$ identified through the Born rule \eqref{eq:Born},
 we now turn to the representation of a scalar supply rate of targets by a complex pair.

In general, $|\bfx_1 + \bfx_2|^2 \ne |\bfx_1|^2 + |\bfx_2|^2$ so that $p$ itself is not directly additive.
It is the mean $\langle p\rangle$ that is observable as an additive long-term average rate,
 and we see that $p\,d\!t$ should be interpreted as the small \emph{probability of a target being present} in small time $d\!t$.

The natural probability assignment for a complex variable $\bfx = x_1 + ix_2$, constrained in modulus by variance $r$ but unconstrained in phase, is Gaussian
\begin{equation} \label{eq:Gauss}
	\Pr(x_1,x_2) =  \frac{1}{\pi r} \exp\!\left(- \frac{x_1^2 + x_2^2}{r} \right)
\end{equation}
and this form is infinitely divisible so (as we would prefer) can be partitioned arbitrarily deeply.
The distribution for $p=|\bfx|^2$ follows as
\begin{equation}
	\Pr(p) =  \frac{1}{r} \exp\!\left(- \frac{p}{r} \right)
\end{equation}
which is exponential.
Its mean is $\langle p\rangle =  r$ so that \eqref{eq:Gauss} is the prior distribution for the pair $\bfx$ controlling a Poisson process of rate $r$.
$$
	\bfx\ \ -\!\!\!-\!\!\!\longrightarrow \ \ p=|\bfx|^2 \ \ -\!\!\!-\!\!\!\longrightarrow \ \ r=\langle p\rangle
$$

It is immaterial whether the analysis describes a single object occasionally entering the target state or whether separate target objects are encountered.
The formalism is the same for either.

We now have the behaviour of pairs (Feynman rules) and their relationship to scalar valuation (Born rule), along with the probabilistic ``Bayesian'' interpretation needed for rational inference.
The foundation is logically complete, and we now develop it into standard quantum formalism.

\subsection{Hilbert space}

We choose to introduce quantum calculus through the notional substrate of $n$ a-priori-equivalent  base states as in (\thetableone),
  for convenience all supplied at the same unit rate.
It is helpful to keep track of combination and partition by inventing orthonormal base vectors $\ket{\bfe_k}$ for these  base states  $k=1,2,\dots,n$.
A sample object $\bfx$ can then be expressed in ``bra-ket'' quantum physics notation as a complex ``amplitude vector'' in Hilbert space.
\begin{equation}
	\ket{\bfx} = \sum_{k=1}^n x_k \ket{\bfe_k}
\end{equation}

With nothing at first known about the independent component amplitudes other than their common unit supply rate,
 their prior distribution is distributed according to the Gaussian multi-dimensional version of \eqref{eq:Gauss}.
\begin{equation} \label{eq:multiGauss}
	\Pr(\bfx_1, \bfx_2,\dots,\bfx_N) = \pi^{-n}  \exp\left( - \sum_{k=1}^n|\bfx_k|^2 \right)
\end{equation}
The supply rate, being the mean of $|\bfx|^2 = \braket{\bfx}{\bfx}$, is $n$, in keeping with unit supply of $n$ components.

According to the pair sum rule \eqref{eq:psum}, composite states covering  base states  from a selected set $S$ have amplitudes
\begin{equation}
	\bfX_S = \sum_{k\in S} \bfx_k
\end{equation}
which are themselves complex Gaussian whose variance is the size (cardinality) of $S$.
Size 1 (a single  base state) is called a ``pure'' state and larger sizes are called ``mixed'',
 maximal mixing with all  base states  included being the sample object itself. 
State $S$ can be extracted from the sample objects by applying the selection operator (in mathematics, a projection)
\begin{equation}
	\bfP_S = \sum_{k\in S} \ket{\bfe_k} \bra{\bfe_k} \qquad\text{so that}\qquad \ket{\bfX_S} = \bfP_S \ket{\bfx}
\end{equation}
Selection separates objects according to some defined property, and is implemented in such devices as diffraction screens and Stern-Gerlach experiments.

The distribution \eqref{eq:multiGauss} of $\bfx$ being spherically symmetric, orthonormal base vectors $\ket{\bfe}$ can be rotated arbitrarily,
 so that a suitable selection $\bfP$ and its complement $\bfQ$ (with $\bfP+\bfQ=\bfI$, the identity) can split the original Hilbert space into any desired subspace and its complement,
 and such partitioning can be carried out to any depth.

\subsection{Single object}

When probing has selected ``one object'' which we then know exists, we specify $|\bfx|^2=1$.
That confines $\bfx$ to the unit Hilbert sphere
\begin{equation}
	\Pr(\bfx_1, \bfx_2,\dots,\bfx_N \mid\text{1 object}) = \frac{\Gamma(\frac{n}{2})}{\pi^{n/2}}\, \delta\Big( 1 - \sum_{k=1}^n |\bfx_k|^2 \Big)
\end{equation}
Component quantities are no longer supply rates but become probabilities of the object being found in individual  base states.
This normalisation constraint would continue to be obeyed no matter how deeply the object is partitioned and recombined.
In this context of one object, the vector $\ket{\bfx}$ is conventionally termed the \emph{wave function} and given the symbol $\ket{\psi}$.

However, we suggest a policy of encoding magnitude within the pairs $\bfx$ themselves.
After all, rates are additive indefinitely, while probabilities are required (rather awkwardly) to sum to unity.
And rates are closer to laboratory practice.
Theorists tend to discuss states while experimentalists provide rates.

\section{Commentary}
``\emph{Now the essential content of both statistical mechanics and communication theory, of course, does not lie in the equations; it lies in the ideas that lead to those equations.}'' --- Edwin T. Jaynes \cite[p.4]{Jaynes:1959}
\medskip

We have presented a unified derivation of summation in measure theory, multiplication in probability theory, and complex numbers in quantum theory.
This minimal foundation is very simple --- just check the boxes --- and should be accessible to neophyte students as well as experienced researchers.
$$
{\scriptsize
\begin{tabular}{|l|c|c|c|}
\hline
				\hfill CHECKLIST					& Measure 	& Probability 	& Quantum	\\
\hline \hline	
	Dimension\ ? 									& scalar		& scalar		& pair		\\
\hline
\phantom{Right-distributive $\circ$}\llap{\it Combination}		& 		 	& 		 	& 			\\
	\rlap{Closure of}\phantom{Commutative}	$\oplus$ 		& \checkmark 	& \checkmark 	& \checkmark	\\
	        Commutative 					$\oplus$		& \checkmark 	& \checkmark 	& \checkmark	\\
	\rlap{Associative}\phantom{Commutative}	$\oplus$		& \checkmark 	& \checkmark 	& \checkmark	\\
	\rlap{Limitless}\phantom{Commutative}	$\oplus$ 		& \checkmark 	& \checkmark 	& \checkmark	\\
\hline
\phantom{Right-distributive $\circ$}\llap{\it Chaining}			& 		 	& 		 	& 			\\
	\rlap{Closure of}\phantom{Right-distributive}	$\circ$ 	& 		 	& \checkmark 	& \checkmark	\\
	        Right-distributive					$\circ$	& 		 	& \checkmark 	& \checkmark	\\
	\rlap{Associative}\phantom{Right-distributive} 	$\circ$ 	& 		 	& 		 	& \checkmark	\\
\hline
\end{tabular}
}
$$
In response to the Rovelli quote \cite{Rovelli:FQXI2013} we started with,
 quantum mechanics may cease to look puzzling now that we have derived the formalism of the theory from a set of simple assertions about the world.
Measurement involves no dubious ``collapse of the wave function''.
We only need ordinary probabilistic reasoning, in which our partial knowledge of an object is modified when it interacts with a probe, and then modulated again if we later choose to retrieve and interrogate the probe.

The formalism is straightforward.
A target --- equivalently an object in a particular target state --- is represented by a number pair.
This turns out to be a single complex number whose squared amplitude represents the Poisson supply rate (the Born rule) which becomes observable through interaction with probes.

As a stream of targets is sent along paths which may split and merge,
 the number pairs evolve through the (Feynman) sum and product rules required by the symmetries of partition and combination.
The standard Hilbert-space structure for multi-state objects follows, and specification from supply rates to single objects is immediate.

We make no assumption that cannot be checked in the lab.
We recommend that as a good strategic principle, because assumptions that cannot be checked are thereby divorced from practical impact,
in which case they become a peculiar and questionable part of scientific inquiry.

If such assumption is truly needed, then it has practical impact after all because its denial would alter experimental results, which is self-contradictory.
If it's not needed, then requiring it would be regrettable.
Specifically, we make no assumption involving infinity or the infinitesimal.
Any general theory must apply to special cases, including simple ones, and it happens that simple examples are sufficient to eliminate all but the one calculus.

In the last couple of decades there has been an effort to reformulate and reconstruct the quantum formalism based on probability theory
 \cite{Youssef:1994, Caves+Fuchs+Schack:2002,Bub:2007,Fuchs+Mermin+Schack:2013}
 and information theory \cite{Timpson:2013,Rovelli:1996,Reginatto:1998,Zeilinger:1999,Fuchs:2002,Clifton+Bub+Halvorson:2003,Goyal:2008,Brukner:2009,Wootters:2013}.
Yet we find that the similarities between the quantum rules and probability/information theory are not due to the fact that one derives from the other,
 but rather that they both derive from common principles.
We share much of the interpretive aspect of quantum Bayesianism (QBism) \cite{Fuchs+Mermin+Schack:2013}.
For us, though, Bayes comes first.

Other than that, symmetries are silent on interpretation and we are content with any outlook that conforms.
The various interpretations of quantum theory (Copenhagen, many worlds, \dots) all use the same calculus and make the same predictions,
 so the choice reduces to free personal preference.

Note that our derivation of the quantum formalism cannot be undermined by any alternative interpretation or supposed generalisation of probability,
 or by some differing assumptions there \cite{Dupre+Tipler:2009,Terenin+Draper:2015}, which might be thought to open the possibility of conflict.
Symmetries are silent on interpretation, and we need only the one common foundation to support the whole edifice.

Our symmetries are necessary and sufficient, but we do not exclude using similarly verifiable assumptions as sufficient foundation \cite{Knuth+Skilling:2012}.
But, as a matter of logic, any alteration to measure (which has not been seriously proposed) or to probability (which has often been proposed) must conflict with our symmetries,
 and thereby with implementations of our quantum thought-experiments.
For, it must be acknowledged, quantum theory works.
So does probability.
And the two are entirely mutually consistent.

\section{Acknowledgements}

We thank those many colleagues who have guided the evolution of our thought over the past quarter-century,
particularly Ariel Caticha, Seth Chaiken, Keith Earle, Anton Garrett, Steve Gull, and Oleg Lunin.
We also thank Andrei Khrennikov, Julio Stern, and Federico Holik for invitations to present our efforts leading up to this manuscript.
K.H.K. also thanks the Foundational Questions Institute (FQXi) and those who have worked to support the FQXi essay contests,
for providing an opportunity for researchers to explore their thoughts and ideas on foundational topics \cite{Knuth:FQXI2015}.
Yet our greatest debt is to Edwin Jaynes who kept faith with rational inference through many dark years, and to whose memory we respectfully dedicate this work.

The authors contributed equally to this work.
The authors declare no competing financial interests.



\begin{thebibliography}{[10]}

\othercit
\bibitem{Rovelli:FQXI2013}
 \textsc{C.~Rovelli},
Relative information at the foundation of physics, 2013,
``It from Bit or Bit from It?'' FQXi 2013 Essay Contest (Second Prize)\\
http://www.fqxi.org/community/forum/topic/1816).


\othercit
\bibitem{Caticha:2009:rational-belief}
 \textsc{A.~Caticha},
Quantifying rational belief,
 in: Bayesian Inference and Maximum Entropy Methods in Science and Engineering,
  Oxford MS, USA 2009, edited by P.\,M. Goggans and C.\,Y. Chan,  (2009),
  pp.\,60--68.


\bibitem{Cox:1946}
 \textsc{R.\,T. Cox}, \jr{Am. J. Physics} \textbf{14}, 1--13 (1946).


\bibitem{Caticha:1998}
 \textsc{A.~Caticha}, \jr{Phys. Rev. A} \textbf{57}(3), 1572--1582 (1998).


\bibitem{GKS:PRA}
 \textsc{P.~Goyal},  \textsc{K.\,H. Knuth}  and  \textsc{J.~Skilling}, \jr{Phys. Rev. A} \textbf{81}, 022109 (2010),
(arXiv:0907.0909 [quant-ph]).


\bibitem{GK:Symmetry}
 \textsc{P.~Goyal} and  \textsc{K.\,H. Knuth}, \jr{Symmetry} \textbf{3}(2),
  171--206 (2011).


\bibitem{Knuth+Skilling:2012}
 \textsc{K.\,H. Knuth} and  \textsc{J.~Skilling}, \jr{Axioms} \textbf{1}(1),
  38--73 (2012).


\bibitem{Feynman:1948}
 \textsc{R.\,P. Feynman}, \jr{Rev. Mod. Phys.} \textbf{20}(2), 367--387 (1948).


\bibitem{Born:1926}
 \textsc{M.~Born}, \jr{Zeit fur Phys} \textbf{38}, 803 (1926).


\othercit
\bibitem{vonNeumann:1996(1955)}
 \textsc{J.~von Neumann},
Mathematical foundations of quantum mechanics, 12 edition (Princeton University
  Press, 1996).


\bibitem{Hamming:1980}
 \textsc{R.\,W. Hamming}, \jr{American Mathematical Monthly} pp.\,81--90 (1980).


\othercit
\bibitem{Knuth:FQXI2015}
 \textsc{K.\,H. Knuth},
The deeper roles of mathematics in physical laws,
 in: Trick or Truth: the Mysterious Connection between Physics and Mathematics,
  edited by A.~Aguirre, B.~Foster,  and Z.~Merali, Springer Frontiers
  Collection (Springer-Verlag, Heidelberg, 2016),  pp.\,77--90,
FQXi 2015 Essay Contest (Third Prize),(arXiv:1504.06686 [math.HO]).


\othercit
\bibitem{Boole:1854}
 \textsc{G.~Boole},
An Investigation of the Laws of Thought (Macmillan, London, 1854).


\othercit
\bibitem{Knuth:measuring}
 \textsc{K.\,H. Knuth},
Measuring on lattices,
 in: Bayesian Inference and Maximum Entropy Methods in Science and Engineering,
  Oxford, MS, USA, 2009, edited by P.~Goggans and C.\,Y. Chan, AIP Conf. Proc.
  1193 (AIP, New York, 2009),  pp.\,132--144,
(arXiv:0909.3684v1 [math.GM]).


\othercit
\bibitem{Knuth:Annalen-der-Physik}
 \textsc{K.\,H. Knuth},
Lattices and their consistent quantification,
\jr{Annalen der Physik}, 1700370 (2018),
(arXiv:1711.07358 [cs.LO]).


\othercit
\bibitem{Laplace:1812}
 \textsc{P.\,S. Laplace},
Th\'{e}orie {A}nalytique des {P}robabilit\'{e}s (Courcier Imprimeur, Paris,
  1812).


\othercit
\bibitem{Kolmogorov:1950}
 \textsc{A.\,N. Kolmogorov},
Foundations of the Theory of Probability (Chelsea, New York, NY, USA, 1950),
English translation and reprinting of Kolmogorov, A., 1933, Grundbegriffe der
  Wahrscheinlichkeitsrechnung (Springer, Berlin).


\bibitem{DeFinetti:1931}
 \textsc{B.~de~Finetti}, \jr{Erkenntnis} \textbf{31}(2), 169--223 (1989),
{E}nglish translation and reprinting of de {F}inetti, {B}.: 1931c,
  `Probabilismo', Logos (Napoli), pp. 163-219.


\othercit
\bibitem{Jaynes:Book}
 \textsc{E.\,T. Jaynes},
Probability Theory: The Logic of Science (Cambridge Univ. Press, Cambridge,
  2003).


\othercit
\bibitem{Weinberg:1992:dreams}
 \textsc{S.~Weinberg},
Dreams of a final theory (Vintage, 1992).


\bibitem{Tikochinsky:1988}
 \textsc{Y.~Tikochinsky}, \jr{Int. J. Theor. Phys.} \textbf{27}(5), 543--549
  (1988).


\bibitem{Bell:1966}
 \textsc{J.\,S. Bell}, \jr{Rev. Mod. Phys.} \textbf{38}, 447--452 (1966).


\bibitem{Kochen+Specker:1967}
 \textsc{S.~Kochen} and  \textsc{E.\,P. Specker}, \jr{J. Math. Mech.}
  \textbf{17}, 59--87 (1967).


\othercit
\bibitem{Jaynes:1959}
 \textsc{E.\,T. Jaynes},
Probability Theory in Science and Engineering, Colloquium Lectures in Pure and
  Applied Science,  Vol.\,4 (Socony-Mobil Oil Company, Inc., Dallas, TX, USA., 1959).\\
  http://bayes.wustl.edu/etj/articles/mobil.pdf


\bibitem{Youssef:1994}
 \textsc{S.~Youssef}, \jr{Mod. Phys. Lett. A} \textbf{9}, 2571--2586 (1994).


\bibitem{Caves+Fuchs+Schack:2002}
 \textsc{C.\,M. Caves},  \textsc{C.\,A. Fuchs}  and  \textsc{R.~Schack}, \jr{Phys. Rev. A} \textbf{65}(2), 022305 (2002).


\bibitem{Bub:2007}
 \textsc{J.~Bub}, \jr{Studies in History and Philosophy of Science Part B:
  Studies in History and Philosophy of Modern Physics} \textbf{38}(2), 232--254
  (2007).


\othercit
\bibitem{Fuchs+Mermin+Schack:2013}
 \textsc{C.\,A. Fuchs},  \textsc{N.\,D. Mermin}  and  \textsc{R.~Schack},
An introduction to {QB}ism with an application to the locality of quantum
  mechanics, 2013,
(arXiv:1311.5253 [quant-ph]).


\othercit
\bibitem{Timpson:2013}
 \textsc{C.\,G. Timpson},
Quantum information theory and the foundations of quantum mechanics (OUP
  Oxford, 2013).


\bibitem{Rovelli:1996}
 \textsc{C.~Rovelli}, \jr{Int. J. Theor. Phys.} \textbf{35}(8), 1637--1678
  (1996).


\bibitem{Reginatto:1998}
 \textsc{M.~Reginatto}, \jr{Phys. Rev. A} \textbf{58}(3), 1775 (1998).


\bibitem{Zeilinger:1999}
 \textsc{A.~Zeilinger}, \jr{Foundations of Physics} \textbf{29}(4), 631--643
  (1999).


\othercit
\bibitem{Fuchs:2002}
 \textsc{C.\,A. Fuchs},
Quantum mechanics as quantum information (and only a little more), 2002,
(arXiv:quant-ph/0205039).


\bibitem{Clifton+Bub+Halvorson:2003}
 \textsc{R.~Clifton},  \textsc{J.~Bub}  and  \textsc{H.~Halvorson}, \jr{Foundations of Physics} \textbf{33}(11), 1561--1591 (2003).


\bibitem{Goyal:2008}
 \textsc{P.~Goyal}, \jr{Phys. Rev. A} \textbf{78}(5), 052120 (2008).


\bibitem{Brukner:2009}
 \textsc{{\v{C}}.~Brukner} and  \textsc{A.~Zeilinger}, \jr{Foundations of
  Physics} \textbf{39}(7), 677--689 (2009).


\bibitem{Wootters:2013}
 \textsc{W.\,K. Wootters}, \jr{Entropy} \textbf{15}(8), 3130--3147 (2013).


\bibitem{Dupre+Tipler:2009}
 \textsc{M.\,J. Dupr{\'e}} and  \textsc{F.\,J. Tipler}, \jr{Bayesian Analysis}
  \textbf{4}(3), 599--606 (2009).


\othercit
\bibitem{Terenin+Draper:2015}
 \textsc{A.~Terenin} and  \textsc{D.~Draper},
Rigorizing and extending the {C}ox-{J}aynes derivation of probability:
  Implications for statistical practice, 2015,
arXiv preprint arXiv:1507.06597.


\end{thebibliography}

\hyphenation{Post-Script Sprin-ger}
\providecommand{\WileyBibTextsc}{}
\let\textsc\WileyBibTextsc
\providecommand{\othercit}{}
\providecommand{\jr}[1]{#1}
\providecommand{\etal}{~et~al.}


%
%
%
%
%
%
%
\end{document}